\documentclass[12pt]{article}
\usepackage{color}
\usepackage{extarrows}
\usepackage{cases}
\usepackage{bbding}
\usepackage{latexsym}
\usepackage{amsmath}
\usepackage{amsfonts}
\usepackage{amssymb}
\usepackage{psfrag}
\usepackage{graphicx}
\graphicspath{{plots/}}
\usepackage{epsfig}
\usepackage{psfrag,epsf}
\usepackage{epsfig}
\usepackage{lscape}
\usepackage{rotating}
\usepackage{bm}
\usepackage{algorithm}
\usepackage{algorithmic}
\usepackage{float}
\usepackage{booktabs}
\usepackage{bm}
\usepackage{comment}
\usepackage{footnote}

\newtheorem{theorem}{Theorem}

\newtheorem{proposition}{Proposition}

\newtheorem{remark}{Remark}

\def\5n{\negthinspace \negthinspace \negthinspace \negthinspace \negthinspace }
\def\4n{\negthinspace \negthinspace \negthinspace \negthinspace }
\def\3n{\negthinspace \negthinspace \negthinspace }
\def\2n{\negthinspace \negthinspace }

\def\0{\mathbf{0}}
\def\1{\mathbf{1}}

\def\5n{\negthinspace \negthinspace \negthinspace \negthinspace \negthinspace }
\def\4n{\negthinspace \negthinspace \negthinspace \negthinspace }
\def\3n{\negthinspace \negthinspace \negthinspace }
\def\2n{\negthinspace \negthinspace }

\def\0{\mathbf{0}}
\def\1{\mathbf{1}}

\begin{document}
\begin{titlepage}
\title{Combining case-control studies for identifiability and efficiency improvement in logistic regression
	}
\author{ Wenlu Tang, Yuanyuan Lin, Linlin Dai and Kani Chen }
\date{}
\maketitle
\thispagestyle{empty}

\footnotetext[1]{
Wenlu Tang (E-mail: wenlu.tang@polyu.edu.hk) is Research Assistant Professor, Department of Applied Mathematics, The Hong Kong Polytechnic University, Hong Kong, China.
 Yuanyuan Lin (E-mail: ylin@sta.cuhk.edu.hk) is Associate Professor,
Department of Statistics, The Chinese University of Hong Kong, Hong Kong, China.
Linlin Dai (E-mail: ldaiab@swufe.edu.cn) is Assistant Professor, Center of Statistical Research, School of Statistics, Southwestern University of Finance and Economics, Chengdu, Sichuan, China.
Kani Chen (E-mail: makchen@ust.hk) is Professor, Hong Kong University of Science and Technology, Clear Water Bay, Kowloon, Hong Kong.
}

\noindent {\bf Abstract}:
 Can two separate case-control studies, one about Hepatitis disease and the other
 about Fibrosis, for example, be combined together?
 It would be hugely beneficial if two or more separately conducted case-control
 studies, even for entirely irrelevant purposes,  can be merged together with a unified analysis that produce better statistical
 properties, e.g., more accurate estimation of parameters. In this paper, we show
 that, when using the popular  logistic regression model, the  combined/integrative analysis
 produces more accurate   estimation
 of the slope parameters  than the single case-control study.
 It is known that,  in a single logistic case-control study,
 the intercept is not identifiable, contrary to prospective studies. In combined   case-control	studies, however,  the intercepts are proved to be
 identifiable under mild conditions.
 The resulting maximum likelihood estimates of the intercepts and slopes are	proved to be consistent and asymptotically normal, with asymptotic variances achieving the semiparametric efficiency lower bound. 	

\vspace{0.2in}

\end{titlepage}

\section{Introduction}
Logistic regression (Cox, 1958) is probably the most fundamental  statistical tool to model categorical dependent variable.
Let  $Y$ be a categorical response variable
and let $X$ be an observable $d$-dimensional vector of covariates.
Without loss of generality, we focus on binary  response typically coded as $0/1$.
A binary logistic regression model assumes
\begin{eqnarray}
	\label{blg}
	P(Y=1|X=x) = 1-P(Y=0|X=x)= \frac{e^{\alpha+\beta^\top x} }{ 1+ e^{\alpha+\beta^\top x}},
\end{eqnarray}
where $\beta\in \mathbb{R}^d$ is the slope parameter and $\alpha\in \mathbb{R}$ is the intercept.
The logistic regression is
often regarded as a special case of the generalized linear models, which
has been extremely popular
in biomedical science  for the study of the effect
of certain exposure to possible disease or hazards. The response is thus often binary, e.g,  $Y=0$ represents the controls (non-disease) and $Y=1$
represents the cases (disease), or there may be more than two categories of $Y$.
In  prospective studies, a sample of subjects or individuals is followed and their respective $Y$ are recorded.
For model (\ref{blg}) under prospective studies,  the samples are simple random sampling from the underlying population. Thus
$\alpha, \beta$ can be consistently estimated by the maximum likelihood estimation (MLE).
However, many diseases  are often fairly rare such that $P(Y=0|X=x)$
is close to 1. Large studies may produce very few diseased cases and thus very little information about the disease of interest.

Case-control study is a primary tool for the study of existing factors related to rare disease incidence,
by taking samples separately from the case population and the control population, when the case population and the control population
are clearly separated, e.g, through disease registry records.
In econometrics, people are interested in the relationship of the covariates and the choices made by individual, e.g,
the choice of transportation from Shanghai to Hong Kong. It would be easier and cheaper to take samples of individuals
from the Hong Kong international airport, West Kowloon High-speed Train Station and Hung Hom train station, which is
a choice-based sampling data (Manski and McFadden, 1981),  than to
take a single sample from the entire population.
Generally speaking,  according to Lawless(1997), sampling schemes that depend on the value of the outcome/response are called
response-selective to response-biased sampling, among which case-control sampling is
the most typical one; see  Manski and  Lerman (1977), Prentice and  Pyke (1979),
Cosslet (1981), Scott and  Wild (1986, 1997), Manski (1993),  Chen and Lo (1999),	Chen (2001),  Fithian and Hastie (2014),  Liu {\it et al.} (2014)  and Chen {\it et al.} (2017) etc.
Another popular sampling design in epidemiology is the case-cohort sampling.
The estimation and inference of case-cohort data with Cox's proportional hazards model,
transformation models and other semiparametric models are studied by Prentice (1986), 	Lu and Tsiatis (2006), Cai and Zeng (2004, 2007), Liu {\it et al.} (2010),  Zeng and Lin (2014), Ni {\it et al.} (2016),
Tao, Zeng and Lin (2017), etc.
Unified estimations for semiparametric linear transformation models, the accelerated failure time model and quantile regression
under general biased sampling schemes were studied by Kim, Lu, Sit and Ying (2013), Kim, Sit and Ying (2016) and Xu {\it et al.} (2017).
In fact, earlier work on nonparametric inference were developed for recovering the distribution function
in two-sample problem in the presence of selection bias under known selection bias weight function; see Vardi (1982, 1985), Qin (1993) and Qin and Zhang (1997).
When the population case percentage is known, large sample theory of the nonparametric maximum likelihood estimate for
semiparametric biased sampling data was established by Gilbert (2000).
A comprehensive discussion on
biased sampling and associated parameter problems  can be found in Qin (2017).

For logistic regression under  case-control sampling, a remarkable result is that,
the prospective estimating equation derived from the MLE is valid for a consistent estimate of the slope parameter, except the intercept term (Prentice and Pyke, 1979). The beauty of this method is its simplicity and ease of inference.
We first point out that for model (\ref{blg}) under a single case-control study as in the literature,
the intercept $\alpha$ and $f(\cdot)$ are not  identifiable, where $f(\cdot)$ is the probability density function of the covariate $X$ in the population.
The rationale behind is that, with a single case-control study, the score function of $\alpha$ lies in the linear space spanned by the score function of
$f(\cdot)$. This is discussed in details in Remark 2 in section 2.1.	 
As a result, contrary to prospective studies,  the intercept cannot be consistently estimated in a single case-control study.
Nonetheless,
a consistent estimate of the intercept  is important in many machine learning problems such as  image classification or pattern recognition,  and it is crucial for better understanding of the incidence of the disease, when prior  knowledge of the event/disease is not available or not reliable.
Despite some nice theory on the estimation of $\beta$ is developed under various circumstances,
no specific construction  of consistent estimation of $\alpha$
in case-control logistic regression is available in the literature. 
Moreover, the consistent  estimate for $\beta$ in single case-control study is not semiparametric efficient.

In this paper, we propose an efficient estimation for the intercepts and slope parameters of the logistic regression models under multiple/distributed case-control studies. With $K$ case-control studies
of different purposes from the same population or
collected
from $K$ heterogeneous subpopulations,  the data can be pooled together for integrative analysis.
The intercepts and slope parameters in the logistic regression models can vary across different studies/subpopulations. By combining $K$ case-control studies, surprisingly,
the intercepts  become identifiable under mild conditions (see Proposition 1 in section 2.1), as the score function of each intercept term no longer lies in the linear space spanned by the score of $f(\cdot)$ (see section 2.2 for detailed discussions);
most importantly, the resulting estimates of the slope parameters are proved to be semiparametric efficient, with asymptotic  variance smaller than the counterparts in single  case-control study (see Remark 4 in section 2.2 and Remark 5  in section 2.4).
The proposed estimation is based on the  maximisation of the nonparametric likelihood function
of the  integrative  data.
The resulting estimates for the intercepts and slope parameters are shown to be consistent, asymptotically normal and asymptotically
efficient.   	An iterative algorithm is employed to compute the maximum likelihood estimates  numerically.

\section{Theory and Methods}

Without loss of generality, we
focus on binary outcome. Suppose that there are $K$ independent studies that might be of different purposes, or there are $K$ heterogeneous subpopulations.
For $k=1,\ldots, K$,
the $d$-dimenisonal explanatory variables $X$ and the binary response $Y_k$  satisfy
\begin{eqnarray}\label{klg}
	P(Y_k=1| {X} = x)= \phi(\alpha_k+ \beta_k^\top  x)\equiv \phi_k(\theta, x), \hskip 1cm k=1,\ldots, K,
\end{eqnarray}
where $\phi(x)= e^x/(1+e^x)$ and $\theta = (\alpha_1,\ldots,\alpha_K, \beta_1^\top, \ldots, \beta_K^\top)^\top$  of dimension  $p=K+Kd$. Let $f(\cdot)$ and $F(\cdot)$  be the density function and the
cumulative distribution of $X$ respectively. We assume  that $f(\cdot)$ and $F(\cdot)$ are  the same for $k=1,\ldots, K$ in model (\ref{klg}).

For each $k=1,\ldots, K$, let $f_{k1}(\cdot)$ be the density function of the case population, i.e, the conditional distribution of $X$
given $Y_k=1$, and $f_{k0}(\cdot)$ be that of the control population, i.e, the conditional distribution of $X$
given $Y_k=0$.
Suppose that the $k$-th case-control study is conducted by
taking a random sample of $n_{k1}$ cases from its case population, and a random sample of $n_{k0}$ controls from its control population,
denoted by $\mathcal{D}_k = \{(y_{ik}, x_{ik})\}_{i=1}^{n_k}$.
Note that $n_{k1}$ and $n_{k0}$ are pre-specified in case-control studies.  Let $n_k=n_{k1}+ n_{k0}$ and $N=\sum_{k=1}^{K}n_k$.
Thus,  the pooled data are $\mathcal{D} = \{\mathcal{D}_k$, $k=1,\ldots,K\}$ consisting of $K$ case-control studies.
Under model (\ref{klg}),  by the Bayes' formula, for $ k=1,\cdots,K,$
\begin{align}
	\label{bayes}
	f_{k1}(x) = \frac{\phi_k(\theta, x)f( x)}{c_{k}(\theta)}, \ \ \ f_{k0}( x) = \frac{\{1-\phi_k(\theta, x)\}f(x)}{1-c_{k}(\theta)},
\end{align}
where
$c_{k}(\theta)=\int_{-\infty}^{+\infty}\phi_{k}(\theta, x)f(x)d  x$ is the population percentage of cases for the $k$-th study or the $k$-th subpopulation.  For notational simplicity,  we simply write $c_{k}(\theta)$ as $c_k$, $k=1, \ldots, K$.
Let $\beta_{k, 0}$
and $\alpha_{k,0}$ be the true values of $\beta_k$ and $\alpha_k$, $k=1,\ldots, K$, respectively.

\subsection{Identifiability}
Throughout the paper, parameters including $f(\cdot)$ are not identifiable if two different choices have the same resulting distribution of random variables. 	In classical logistic regression model (\ref{blg}) under single  case-control study (Scott and Wild, 1986), that is $K=1$ in model (\ref{klg}),
for any two  values $x$ and $x_0$ in the support of $X$,
the logarithmic odds ratio (OR)  is
\begin{equation}\label{OR}
	\log \{\mathrm{OR}(x)\}=\log \left\{\frac{ f(x|Y=1)/f(x_0|Y=1)}{ f(x|Y=0)/f(x_0|Y=0)}\right\}=\beta^\top (x-x_0),
\end{equation}
where $f(\cdot|Y=1)$ is $f_1(\cdot)$,  the conditional density of $X$ in the case population  defined earlier.
Heuristically, independent samples from the case population and control distribution are observed separately in case-control studies, making $\beta$  identifiable according to (\ref{OR});
but $\alpha$ is not involved in (\ref{OR}) and thus it is not identifiable under  single case-control study.

With multiple/distributed case-control studies,  however, the data can be pooled together
for unified analysis.  Under regularity conditions (C1)-(C3) given in section 2.4,   the idea of combining data enables us to
consistently estimate the intercepts.  We discuss the identifiability of $\theta$ under multiple case-control studies in the following proposition. 

\noindent	\begin{proposition} \ 
	\begin{itemize}     
		\item[(a)] The slope parameters $\beta_k$, $k=1,..., K$, are always identifiable.
		\item[(b)] For some  $k \leq K$, if $\beta_k \not=0$, then
		$\alpha_k$ is  identifiable if and only if $f(\cdot)$ is identifiable.
		As a result, $f(\cdot)$ is identifiable if and only if
		one of the $\alpha_k$ is  identifiable.
		
		\item[(c)]   If $\beta_1=...=\beta_K$ and $\alpha_1=\alpha_2=\ldots=\alpha_K$, then all $\alpha_k$ are not  identifiable.
		
		\item[(d)] Assume $f$ is continuous, and $\beta_k$ are not all same.
		Then,  those $\alpha_k$ with $\beta_k=0$ are not identifiable and those with
		$\beta_k \not=0$ are not identifiable.		\end{itemize}
\end{proposition}
Part(a) is consistent with the results in Prentice and Pyke (1979). Part(b)
tells the  relationship between the identifiability of the intercepts and $f(\cdot)$. Part(c) has an important implication that when each separate  study/subpopulation shares the same/common intercept and slope parameters,  the intercept is still not identifiable despite the data combining. In other words,  when a single case-control study is sub-divided 
into multiple smaller studies, combining those smaller case-control studies 
cannot make  the intercept identifiable.		
Part(d) implies that those sub-models with $\beta_k=0$,  their respective $\alpha_k$ is always not identifiable, regardless of the identifiability of $f(\cdot)$. 
For the rest of this paper, we assume that
$\theta = (\alpha_1,\ldots,\alpha_K, \beta_1^\top, \ldots, \beta_K^\top)^\top$ is identifiable.

\subsection{The Score Functions and Maximum Likelihood Estimation }

We next introduce our proposed maximum likelihood estimation.
The likelihood function of $\{\theta, f(\cdot)\}$ under $K$ case-control studies is
\begin{equation}
	\label{lik}
	\mathcal{L}(\theta, f;\mathcal{D})=\prod_{k=1}^{K}\prod_{i=1}^{n_k} f_{k1}( x_{ik})^{y_{ik}}f_{k0}( x_{ik})^{(1-y_{ik})}
\end{equation}
and the log-likelihood function is
\begin{align}
	\label{loglike}
	l( \theta, f;\mathcal{D}) \ = \ &-\sum_{k=1}^{K}\big [n_{k1}\log c_{k}(\theta)+n_{k0}\log \{1-c_{k}(\theta)\}\big ]+\sum_{k=1}^{K}\sum_{i=1}^{n_k}\log f(x_{ik})\nonumber\\
	&+\sum_{k=1}^{K}\sum_{i=1}^{n_k}\big [y_{ik}\log \phi_{k}( \theta , x_{ik})+(1-y_{ik})\log \{1-\phi_{k}(\theta , x_{ik})\}\big].
\end{align}
According to the profile-likelihood method in Zeng and Lin (2006, 2007), the nonparametric component
$f(\cdot)$ in (\ref{loglike}) can be profiled over its observed values.  To avoid uninteresting discussions,
we assume there is no tie in the realizations of $X$. With slight abuse of notation,
we still use  $x_{ik}, i=1,\ldots, n_k, k=1,\ldots, K$ to denote the realizations.
We define the estimator of $F(\cdot)$ as a step function with jumps only at the distinct observed values
$x_{ik}, i=1,\ldots, n_k, k=1,\ldots, K$, that is
$ F\{x_{ik} \} \equiv  p_{ik} $   for $ i=1,\ldots,n_k, k= 1,\ldots,K $ and $0$ otherwise,
where $p_{ik}\geq 0$ is the jump size of $F(\cdot)$ at $x_{ik}$
satisfying $\sum_{k=1}^{K}\sum_{i=1}^{n_k}p_{ik}=1$  and  $\tilde{c}_t(\theta)=\sum_{k=1}^{K}\sum_{i=1}^{n_k}\phi_t(\theta, x_{ik})p_{ik}\equiv \tilde c_t, t=1,\ldots, K.$
Write $ p \equiv \{p_{ik},k=1,\ldots, K, \ i=1, \ldots, n_k\}$.
The log-likelihood function after discretization is
\begin{align}
	\label{profilelike}
	\tilde l(\theta,  p;\mathcal{D})	
	&= -\sum_{t=1}^{K}\big \{n_{t1}\log \tilde c_{t}+n_{t0}\log (1-\tilde c_{t})\big \}+\sum_{k=1}^{K}\sum_{i=1}^{n_k}\log p_{ik}\nonumber\\
	&\qquad \ +\sum_{k=1}^{K}\sum_{i=1}^{n_k}\big [y_{ik}\log \phi_{k}(\theta , x_{ik})+(1-y_{ik})\log \{1-\phi_{k}( \theta , x_{ik})\}\big],
\end{align}
subject to
\begin{eqnarray}
	\label{cnstn}
	&&\sum_{k=1}^{K}\sum_{i=1}^{n_k}p_{ik}=1, \quad p_{ik}\geq 0, \quad  i=1,\ldots,n_k, \quad k=1,\ldots,K, \nonumber \\
	&& \sum_{k=1}^{K}\sum_{i=1}^{n_k}\phi_{t}(\theta, x_{ik})p_{ik}= \tilde{c}_t, \quad t=1,\ldots,K.
\end{eqnarray}
Our proposed maximum likelihood estimator for $\theta$ and $p_{ik}$, denoted by
$\hat\theta$ and $\hat p_{ik}$,  is defined as the maximizer of
$\tilde l_n(\theta, p;\mathcal{D})$ under the constraints  in (\ref{cnstn}). 
Thus,  the nonparametric maximum likelihood estimator
(NPMLE) for
$F(x)$ is given by
$\hat F_n(x)=\sum_{x_{ik}\leq x}\hat p_{ik}$.

\begin{remark}
	Let $\rm{vec}(A)$ be the vectorization of matrix $A$ of order $m \times n$, namely,
	\begin{equation*}
		\mathrm{vec}({A})=[a_{1,1},\ldots,a_{1,n},\ldots, a_{m,1},\ldots,a_{m,n}]^\top,
	\end{equation*}
	where $a_{i,j}$ is the $(i,j)$-th element in $A$.
	The score function of $\theta$, denoted by $S_{\theta}$,  is
	\begin{align}
		\label{score_the}
		S_{\theta}\nonumber \equiv &(S_{\alpha_1},\ldots,S_{\alpha_K},S_{\beta_1}^\top,\ldots,S_{\beta_K}^\top)^\top
		=\frac{\partial l( \theta, f;\mathcal{D})}{\partial \theta} \nonumber \\
		=&-\sum_{t=1}^{K}\left(\frac{n_{t1}}{c_t}-\frac{n_{t0}}{1-c_t}\right)\sum_{k=1}^{K}\sum_{i=1}^{n_k}\phi_t(\theta,x_{ik})\{1-\phi_t(\theta,x_{ik})\}f(x_{ik})
		\binom
		{e_t}{\mathrm{vec}(e_tx_{ik}^\top)}
		\nonumber\\&+\sum_{t=1}^{K}\sum_{i=1}^{n_t}\{y_{it}-\phi_t(\theta,x_{it})\}\binom
		{e_t}{\mathrm{vec}(e_tx_{it}^\top)},
	\end{align}
	where $e_t$ is a $K$-vector with the $t$-th element being $1$ and other elements being $0$. And the score function of $f$, denoted by $S_f[g_1]$, is the partial derivative of $l( \theta, f;\mathcal{D})$ along the direction $(\theta, f_\varepsilon=f+\varepsilon g_1)$ with $\varepsilon$ being a small constant, such that $f_\varepsilon\geq 0$ and
	$$g_1\in  \mathcal{G} \equiv \left\{g\in BV[\mathbb{R}^{d}]: \int g(x)dx=0\right\}.$$
	Here,  $BV[D]$ is a class of functions on domain $D$ with bounded total variation.
	Then,
	\begin{equation*}
		S_f[g_1]=\sum_{k=1}^{K}\sum_{i=1}^{n_k}\left(-\sum_{t=1}^{K}\left[\frac{n_{t1}}{c_t}\phi_t(\theta,x_{ik})+\frac{n_{t0}}{1-c_t}\{1-\phi_t(\theta,x_{ik})\}\right]+\frac{1}{f(x_{ik})}\right)g_1(x_{ik}).
	\end{equation*}
	Given the observations  $x_{ik}$, $i=1,\ldots, n_k$,
	$k=1,\ldots, K$,
	the log-likelihood $l( \theta, f;\mathcal{D})$ can be  approximated by $\tilde l(\theta,  p;\mathcal{D})$; and  the score function of $p_{ik}$, $i=1,\ldots, n_k$, $k=1,\ldots, K$,  based on the profile likelihood $\tilde l(\theta,  p;\mathcal{D})$	is
	\begin{align}\label{score_f}
		S_f^\ast (p_{ik})\equiv -\sum_{t=1}^{K}\left[\frac{n_{t1}}{\tilde c_t}\phi_t(\theta,x_{ik})+\frac{n_{t0}}{1-\tilde c_t}\{1-\phi_t(\theta,x_{ik})\}\right]+\frac{1}{p_{ik}}.
	\end{align}	
	Moreover,  the linear space spanned by  $S_f^\ast (p_{ik})$,  $i=1,\ldots, n_k$, $k=1,\ldots, K$,     is a $N$-dimensional surface on $S_f$ (Bickel {\it et al.} 1993). As a result, the linear space spanned by $S_f$ contains the linear space spanned by  $S_f^\ast$.
	\begin{align*}
		&\sum_{i=1}^{n}S_{f}^\ast(p_i) p_i\{y_i-\phi(\theta,x_i)\}=S_{\alpha}\\
		=&\sum_{i=1}^{n}\left[-\left\{\frac{n_1}{\tilde c}\phi(\theta,x_i)+\frac{n_0}{1-\tilde c}(1-\phi(\theta,x_i))\right\}p_i\{y_i-\phi(\theta,x_i)\}+\{y_i-\phi(\theta,x_i)\}\right],
	\end{align*}
	where the equality holds by invoking (\ref{score_the}).
	This observation reveals that  $S_\alpha$ and $S_f^\ast$ are on the same space when $K=1$; in other words,  $S_\alpha$ is on the linear space spanned by $S_f$ when $K=1$. As a result,  the intercept and the population percentage of cases are not identifiable in  single case-control  logistic regression.
\end{remark}

\begin{remark}
	When $K\ge 2$,  we focus on $K=2$ without loss of generality.
	
	\begin{itemize}	
		\item When $\beta_1\neq \beta_2$, $\beta_1\neq 0$ and $\beta_2\neq 0$,
		the score functions of $\alpha_1$ and $\alpha_2$ are
		\begin{align}\label{rmk1}
			&S_{\alpha_1}=-\left(\frac{n_{11}}{c_1}-\frac{n_{10}}{1-c_1}\right)
			\sum_{i=1}^{n_1}\phi(\alpha_1+\beta_1x_{i1})\{1-\phi(\alpha_1+\beta_1x_{i1})\}+n_{11}-\sum_{i=1}^{n_1}\phi(\alpha_1+\beta_1x_{i1}),\\
			&S_{\alpha_2}=-\left(\frac{n_{21}}{c_2}-\frac{n_{20}}{1-c_2}\right)\sum_{i=1}^{n_2}\phi(\alpha_2+\beta_2x_{i2})\{1-\phi(\alpha_2+\beta_2x_{i2})\}+n_{21}
			-\sum_{i=1}^{n_2}\phi(\alpha_2+\beta_2x_{i2}) \nonumber
		\end{align}
		respectively. Similar to Remark 1,  direct calculations yield that the surface of the score function $S_f$ is
		\begin{equation*}
			S_f(p_{ik})=-\sum_{t=1}^{2}\left[\frac{n_{t1}}{\tilde c_t}\phi(\alpha_t+\beta_tx_{ik})+\frac{n_{t0}}{1-\tilde{c}_t}(1-\phi(\alpha_t+\beta_tx_{ik}))\right]+\frac{1}{p_{ik}}.
		\end{equation*}
		Apparently, the score functions of $\alpha_1$ and $\alpha_2$ are not on any spanned linear space of $S_f$
		when $\beta_1\neq \beta_2$, $\beta_1\neq 0$ and $\beta_2\neq 0$.
		This fact  offers insights that the intercepts  $\alpha_1, \alpha_2$ can be identifiable by combining the two case-control studies, when $\beta_1$ and $\beta_2$ are different and nonzero.
		Thus,  $\alpha_1, \alpha_2, \beta_1, \beta_2$ and $f(\cdot)$ in model (\ref{klg})
		can be  consistently estimated by the proposed maximum likelihood estimation.

		\item	 When $\beta_1\neq 0$ and $\beta_2= 0$,
		the score function of $\alpha_1$ is (\ref{rmk1}) and that of  $\alpha_2$ is
		$$S_{\alpha_2}=-\left(\frac{n_{21}}{c_2}-\frac{n_{20}}{1-c_2}\right)\phi(\alpha_2)\{1-\phi(\alpha_2)\}+n_{21}-n_2\phi(\alpha_2),$$
		indicating that $S_{\alpha_2}$ is  in the linear space spanned by $S_f$,  but  $S_{\alpha_1}$ is not. 
		In other words,   $\alpha_2$ is unidentifiable but $\alpha_1$ is  identifiable. This is consistent with Proposition (1d) in section 2.1.

		\item	Lastly, for the special case that $\beta_{1}=0$ and $\beta_2=0$,
		the score functions of $\alpha_1$ and $\alpha_2$ are
		\begin{align*}
			S_{\alpha_1}&=-\left(\frac{n_{11}}{c_1}-\frac{n_{10}}{1-c_1}\right)\phi(\alpha_1)\{1-\phi(\alpha_1)\}+n_{11}-n_1\phi(\alpha_1),\\
			S_{\alpha_2}&=-\left(\frac{n_{21}}{c_2}-\frac{n_{20}}{1-c_2}\right)\phi(\alpha_2)\{1-\phi(\alpha_2)\}+n_{21}-n_2\phi(\alpha_2)
		\end{align*}
		respectively.
		It is clear that $S_{\alpha_1}$ and $S_{\alpha_2}$ are on the spanned linear space of $S_f$. As a result,  $\alpha_1$ and $\alpha_2$ are unidentifiable when $\beta_{1}=0$ and $\beta_2=0$.
	\end{itemize}	
\end{remark}

\begin{remark}
	We need to emphasize that,  $\beta_k$  in model (\ref{klg}) can be always separately estimated
	by the celebrated estimation in  Scott and Wild (1986) or Qin (2017, pages 210-212) with
	the $k$-th case-control study,  $k=1,\ldots, K$, though $\alpha_k$ is unidentifiable.
	In contrast to the classical separate estimation in the literature,
	our proposed method based on the integrative data is able to identify the intercepts and give more efficient estimation of $\beta_k$, $k=1,\ldots,K$, which achieves the semiparametric efficiency lower bound, as evidenced in the simulation studies.
\end{remark}

\subsection{An Iterative Algorithm }

Direct maximization of (\ref{profilelike})  is challenging and potentially
unstable as the second constrain in (\ref{cnstn}) is intractable.
To circumvent the difficulty,  we first plug in  $ \tilde{c}_t= \sum_{k=1}^{K}\sum_{i=1}^{n_k}\phi_{t}(\theta, x_{ik})p_{ik},$ $t=1,\ldots,K$ into
(\ref{profilelike}),    and consider to maximize a Lagrange function  by incorporating the first constraint of (\ref{cnstn}) as follows
\begin{align}\label{lagrange1}
	&-\sum_{t=1}^{K}\left[n_{t1}\log \sum_{k=1}^{K}\sum_{i=1}^{n_k}\phi_{t}(\theta, x_{ik})p_{ik}+n_{t0}\log \{1- \sum_{k=1}^{K}\sum_{i=1}^{n_k}\phi_{t}(\theta, x_{ik})p_{ik}\}\right]+\sum_{k=1}^{K}\sum_{i=1}^{n_k}\log p_{ik} \nonumber\\
	&+\sum_{k=1}^{K}\sum_{i=1}^{n_k}\left[y_{ik}\log \phi_{k}(\theta , x_{ik})+(1-y_{ik})\log \{1-\phi_{k}( \theta , x_{ik})\} \right]-\lambda\left(1-\sum_{k=1}^{K}\sum_{i=1}^{n_k}p_{ik}\right),
\end{align}
where   $\lambda>0$ is the Lagrange multiplier.   Taking first derivative of (\ref{lagrange1})
with respect to $p_{ik}$ and $\theta$, we obtain
\begin{equation}\label{deriv_p}
	\frac{\partial \tilde l(\theta,\bm p;\mathcal{D})}{\partial p_{ik}}=\frac{1}{p_{ik}}-\sum_{t=1}^{K}\left[\frac{n_{t1}}{\tilde c_t}\phi_t(\theta,x_{ik})+\frac{n_{t0}}{1-\tilde c_t}\{1-\phi_t(\theta,x_{ik})\}\right]-\lambda=0,
\end{equation}
and
{
	\begin{align}\label{deriv_the}
		\frac{\partial \tilde l(\theta,\bm p;\mathcal{D})}{\partial \theta}=&-\sum_{t=1}^{K}\left(\frac{n_{t1}}{\tilde c_t}-\frac{n_{t0}}{1-\tilde c_t}\right)\sum_{k=1}^{K}\sum_{i=1}^{n_k} \phi_t(\theta,x_{ik})\{1-\phi_t(\theta,x_{ik})\}p_{ik}\binom
		{e_t}{\mathrm{vec}(e_tx_{ik}^\top)}  \nonumber\\
		&+\sum_{k=1}^{K}\sum_{i=1}^{n_k}\{y_{ik}-\phi_k(\theta,x_{ik})\}\binom
		{e_k}{\mathrm{vec}(e_kx_{ik}^\top)}=0.
\end{align}}
Next, multiplying $p_{ik}$ on both sides of (\ref{deriv_p}), one can easily verify that
$\lambda=0$ because of  $N=\sum_{t=1}^{K}n_t$.
Consequently,
\begin{equation}\label{p}
	{p_{ik}}=\frac{1}{\sum_{t=1}^{K}\left[\frac{n_{t1}}{\tilde c_t}\phi_t(\theta,x_{ik})+\frac{n_{t0}}{1-\tilde c_t}\{1-\phi_t(\theta,x_{ik})\}\right]}.
\end{equation}
We now devise an iterative algorithm to maximize (\ref{lagrange1}).
Set the initial value of the parameter $\bm p$ as $\bm{p}^{0}$, where $\bm{p}^{(0)} \equiv \{p^{(0)}_{ik},k=1,\ldots, K, \ i=1, \ldots, n_k\}$.  For the $j$-step, given $\bm{p}^{(j)}$,  we update $\theta^{(j)}$  by solving
\begin{align}\label{A4}
	&-\sum_{t=1}^{K}\left\{\frac{n_{t1}}{\hat{c}_t(\theta,\bm{p}^{(j)})}-\frac{n_{t0}}{1-\hat{c}_t(\theta,\bm{p}^{(j)})}\right\}\sum_{k=1}^{K}\sum_{i=1}^{n_k} \phi_t(\theta,x_{ik})\{1-\phi_t(\theta,x_{ik})\}p_{ik}^{(j)}\binom
	{e_t}{\mathrm{vec}(e_tx_{ik}^\top)}\nonumber\\&+\sum_{k=1}^{K}\sum_{i=1}^{n_k}\{y_{ik}-\phi_t(\theta,x_{ik})\}\binom
	{e_k}{\mathrm{vec}(e_kx_{ik}^\top)} =0
\end{align}
for $\theta$, where $\hat{c}_t({\theta},\bm{p}^{(j)})=\sum_{k=1}^{K}\sum_{i=1}^{n_t}\phi_t({\theta},x_{ik}){p}_{ik}^{(j)}$.
Notice that the objective function in (\ref{lagrange1}) is convex in $\theta$, thus the solution to (\ref{A4}) is unique.
Subsequently, given $(\theta^{(j)},\bm{p}^{(j)})$, we update  $\bm{p}^{(j+1)}$ according to  the recursive formula
\begin{equation}\label{A3}
	\hat{p}_{ik}^{(j+1)}=1\big{/}\sum_{t=1}^{K}\left[\frac{n_{t1}}{\hat{c}_t(\theta^{(j)},\bm{p}^{(j)})}\phi_t(\theta^{(j)},x_{ik})+\frac{n_{t0}}{1-\hat{c}_t(\theta^{(j)},\bm{p}^{(j)})}\{1-\phi_t(\theta^{(j)},x_{ik})\}\right]. \end{equation}
We iterate between (\ref{A4}) and (\ref{A3}) until convergence. Denote the resulting estimator for $\theta$ and  $\bm p$ by $\hat{\theta}$ and $\hat{\bm p}$. Thus,
$\hat{F}_n(x)\equiv\sum_{k=1}^{K}\sum_{i=1}^{n_k}\hat{p}_{ik}I(x_{ik}\le x)$.  Moreover, the population percentage of cases $\tilde{c}_t$
is estimated by  $ \sum_{k=1}^{K}\sum_{i=1}^{n_k}\phi_{t}(\hat \theta, x_{ik})\hat p_{ik}$,  $t=1,\ldots, K$.

The iterative algorithm is summarized as follows:
	\begin{algorithm}[H]
	\caption{The iterated algorithm for computing the MLEs }
	\label{alg:A}
	\begin{algorithmic}
		\STATE {Set  an initial value $\bm{p}^{(0)}$.} 
		\REPEAT
		\STATE {\it Step 1.} Given $\bm{p}^{(j)}$, compute $\theta^{(j)}$ by solving (\ref{A4}) with the {\it lgfbs} package in R.
		\STATE {\it Step 2.} Given $(\theta^{(j)},\bm{p}^{(j)})$, update $\bm{p}^{(j+1)}$ according to (\ref{A3}).
		\UNTIL{ $|\theta^{(j+1)}-\theta^{(j)} |  \leq \kappa_1$ and  $|\bm p^{(j+1)}-\bm p^{(j)} |  \leq \kappa_2$.}
	\end{algorithmic}
\end{algorithm}
\noindent   Here $\kappa_1$ and $\kappa_2$ in the stopping criterions are set to be $n^{-6}$ in our numerical studies.

\subsection{Asymptotic properties }
Let $\theta_0\equiv(\alpha_{1,0},\ldots,\alpha_{K,0},\beta_{1,0}^\top, \ldots,\beta_{K,0}^\top)$ and $F_0$ be the true values of $\theta$ and $F$.  Let $ \mathcal{G}=\{g\in BV[\mathbb{R}^d]: |g|\leq 1\}$. Here $BV[D]$ is the set of functions on domain $D$ with bounded total variation. And $\sqrt{N}(\hat{F}_n-F_0)$ can be treated as a linear functional in $L^\infty(\mathcal{G})$, the space of all bounded linear functionals on $\mathcal{G}$. The following regularity conditions are imposed:

\begin{itemize}
	
	\item[(C1)] The density $f_0$  is  continuous with bounded support.
	\item[(C2)]  If $ P( X^\top \bm u=0)=1$ for some constant vector $\bm u$, then $\bm u=\bm 0$.
	\item[(C3)]  The true parameter $\theta_0 \in \mathcal{B}_0$, where $\mathcal{B}_0$ is a compact set.
	
\end{itemize}

Conditions (C1)-(C2) are regularity conditions to ensure the identifiability of the parameters. 
The continuously distributed condition of $X$ is imposed for technical convenience. It can be weaken to accommodate discrete predictors, as long as it is satisfied that 
there exist at least three different values   $x_1, x_2, x_3$ in the support of $X$ such that $\beta_{k,0}^\top x_1$, $\beta_{k, 0}^\top x_2$ and $\beta_{k, 0}^\top x_3$ are also different for those nonzero $\beta_{k,0}$, $k=0,\ldots, K$.  
Condition  (C3) assumes that  $\theta_0$ is an interior point of a compact set.
We next present the asymptotic properties in three theorems.

\begin{theorem}{(Consistency of $\hat{\theta}$ and $\hat{F}_n$)}
	Assume conditions (C1)-(C3) hold. Then,
	if $n_k/N\to \rho_k$ for some constant $\rho_k$,  as $n_k\to \infty$ for all $k=1,\ldots, K$,  		
	\begin{equation*}
		|\hat{\theta}-\theta_0|\to 0 \quad {\rm and } \quad \sup_{\bm x\in \mathbb{R}^d}|\hat{F}_n(\bm x)-F_0(\bm x)|\to 0
	\end{equation*}
	almost surely.
	
\end{theorem}


\begin{theorem}{(Asymptotic normality of $\hat{\theta}$ and $\hat{F}_n$ )}
	Suppose	that conditions (C1)-(C3) hold. Then,  if $n_k/N\to \rho_k$ for some constant $\rho_k$,  as $n_k\to \infty$ for all $k=1,\ldots, K$,  	
	$\sqrt{N}(\hat{\theta}-\theta_0, \hat{F}_n-F_0)$ converges weakly to a zero-mean Gaussian process in the metric space $\mathbb{R}^{Kd+K}\times L^\infty(\mathcal{G})$.  The limiting covariance matrix of $\sqrt{N}(\hat{\theta}-\theta_0)$ attains the semiparametric efficiency bound.
\end{theorem}

\begin{theorem}{(Covariance matrix of $\hat{\theta}$ and $\hat{F}_n$)}
	For any $(v,g_1)\in \mathcal{V}\times \mathcal{G}$, where $\mathcal{V}=\{v\in \mathbb{R}^{Kd+K}:|v|\leq 1 \}$, the asymptotic covariance matrix for $$\sqrt{N}v^\top(\hat{\theta}-\theta_0)+\sqrt{N}\int_X \vec{g}_1(X)d\{\hat{F}_n(X)-F_0(X)\}$$ can be estimated by $(v^\top, \vec{g_1}^\top)I_n^{-1}(v^\top, \vec{g_1}^\top)^\top$, where $NI_n$ is the negative Hessian matrix of the log-likelihood $\tilde l(\theta,\bm p;\mathcal{D})$ with respect to $(\theta,\bm p)$ and $\vec{g_1}=(g_1(x_{11}),\ldots,g_1(x_{nK}))$. By taking $\vec{g}_1=0$, the covariance matrix of $\sqrt{N}(\hat{\theta}-\theta_0) $ can be
	estimated by the upper left $(Kd+K)\times (Kd+K)$ matrix of $I_n^{-1}$.
\end{theorem}

Theorem 1 and Theorem 2 indicate that maximizing the profile log-likelihood (\ref{profilelike})  over $(\theta, \bm p)$ leads to  consistent, asymptotically
normal and semiparametric efficient estimate for $\theta$ under mild conditions. Theorem 3 provides a simple and easy-to-implement variance estimation.

\begin{remark}
	Our proposed estimator is semiparametric efficient under $K$ case-control studies by Theorem 2. To have more insights into the efficiency gain, we consider model (\ref{klg})
	with $K=2$,  and focus on the estimation of   the parameter $\theta_1\equiv (\alpha_1, \beta_1^\top)^\top$  of the first study as a toy example.  Three estimators of $\theta_1$ are considered:
	the maximum likelihood estimator for single case-control study with known $f(\cdot)$, denoted by $\hat{\theta}_1^{(1)}=(\alpha_1^{(1)},
	{\beta_1^{(1)\top}})^\top $;  the maximum likelihood estimator  for single case-control study with unknown $f(\cdot)$ (Scott and Wild, 1986), denoted by $\hat{\theta}_1^{(2)}=(\alpha_1^{(2)}, \beta_1^{(2)\top})^\top   $;  our proposed estimator by combining two case-control studies,  denoted by $\hat{\theta}_1^{(3)}= (\alpha_1^{(3)}, \beta_1^{(3)\top})^\top$.
	
	When $f(\cdot)$ is known in a single case-control study, the score function of ${\theta}_1$ is
	\begin{equation*}
		S^{(1)}(\theta_1)\equiv \{y-\phi(\alpha_1+\beta_1^\top x)\}\binom{1}{x}-E\left[\{y-\phi( \alpha_1+\beta_1^\top x )\}\binom{1}{x}|y \right].
	\end{equation*}
	When $f(\cdot)$ is unknown in a single case-control study,  according to Remark 2,
	the score of $\alpha_1$,  $\phi(\alpha_1+\beta_1^\top x )-E\{\phi(\alpha_1+\beta_1^\top x )|y\}$,  is on the linear space spanned by $S_f$.
	Write $\psi_k\equiv \phi(\alpha_k+\beta_k^\top x )x-E\{\phi(\alpha_k+\beta_k^\top x )x|y\}$, $k=1,2$. Then,
	the score function of $\beta_1$
	\begin{equation*}
		S^{(2)}(\beta_1)\equiv \psi_1-E\psi_1-[H(x)-E\{H(x)|y\}],
	\end{equation*}
	where $H(\cdot)\in \mathcal{G}$ is the projection of $\psi_1-E\psi_1 $ onto the linear space spanned by $S_f$ that minimizes $\mathrm{Var}\{S^{(2)}(\beta_1)\}$.
	Apparently, $\mathrm{Var}\{ S^{(1)}(\beta_1)\}$ is always larger than $\mathrm{Var}\{S^{(2)}(\beta_1)\}$.
	Nonetheless, by combining the two case-control studies, the score function of $\theta_1$ becomes
	\begin{align*}
		S^{(3)}(\theta_1)\equiv &\{\psi_1-E(\psi_1|y)\}-\left[g(x)-E\{g(x)|y\}+b_2\{\psi_2-E(\psi_2|y)\}\right],
	\end{align*}
	where $b_2$ is some constant and $g(\cdot)\in \mathcal{G}$  such that $\mathrm{Var}\{S^{(3)}(\theta_1)\}$ is minimized and  nonnegative definite.
	
	Theoretically speaking, the maximum likelihood estimator $\hat{\theta}_1^{(1)}$ for single case-control study with known $f(\cdot)$ is most efficient
	for estimating $\theta_1$, thus combining another case-control study cannot improve efficiency.  However, when $f(\cdot)$
	is unknown, combining case-control studies would lead to efficiency improvement in the estimation of $\theta_1$.
	For instance,  when the sample size of the second case-control study to be combined is much larger than that of the first one,
	our proposed estimator  $\hat{\theta}_1^{(3)}$
	can be nearly as efficient as $\hat{\theta}_1^{(1)}$ asymptotically, implying $\hat{\beta}_1^{(3)}$ is more efficient than $\hat{\beta}_1^{(2)}$ for estimating
	the slope parameter $\beta_1$. In other words, it can be shown that,
	\begin{align*}
		\lim\limits_{r_1/r_2\to 0}\frac{1}{r_1}\mathrm{Var}\{S^{(1)}(\theta_1)\}&=\lim\limits_{r_1/r_2\to 0}\frac{1}{r_1}\mathrm{Var}\{S^{(3)}( \theta_1 )\} , 
	\end{align*}
	where $r_k=n_k/N$, $k=1, 2$.
	
\end{remark}

\section{Empirical results}
Finite-sample studies are carried out to examine the performance of our proposed method under different scenarios.
We consider binary outcome and generate independent data from the following
models
\[P(Y_k=1| X_k= x) =  \frac{1}{1+e^{-(\alpha_{k}+\beta_{k}^\top x)}}\equiv \phi(\alpha_{k}+\beta_{k}^\top x), \ \ k= 1,2,\ldots, K, \]
where $\beta_k \in \mathbb{R}^d$, $k= 1,\ldots, K$. The $d$-vector predictor $X$ follows the same distribution $F(\cdot)$  across $K$ logistic regression models.
Thus, the true population percentage of cases  in the $k$-th subpopulation can be calculated via $
\label{eprob}
p_k\equiv P(Y_k=1)= \int \phi(\alpha_{k}+\beta_k^\top x)  d F(x)$,  $k=1,...,K$.
The case-control sampling are then conducted by taking samples separately from the case population and the control population from each subpopulation respectively.  Let $q_k=n_{k1}/n_k, k = 1,...,K$, be the case proportion in the $k$-th case-control study.

\subsection{Simulated data: verifying identifiability conditions}
We investigate the identifiability of $K$ intercepts under various settings.
In cases (a1)--(a4),  $X_1$ and $X_2$ are independent standard normal random variables. For simplicity, set $n_1=n_2$, $n_{11}=n_{10}$ and $n_{21}=n_{20}$. The pooled sample size $N=n_1+n_2=500$. The following scenarios (a1)--(a6) are tried for $K=2$ and $d=2$. The results of each case are based on 1000 replications.
\begin{enumerate} 
	\itemsep-3pt
	\item[(a1)] ($\alpha_1\neq\alpha_2$, $\beta_1\neq0$,  $\beta_2\neq0$) Set $\theta_1 = (2,2,3)^\top$, $\theta_2 = (-1,3,2)^\top$; then, $p_1\approx 0.691$ and $p_2\approx 0.402$;
	\item[(a2)] ($\alpha_1=\alpha_2$, $\beta_1\neq0$, $\beta_2\neq0$) Set $\theta_1 = (2,2,3)^\top$, $\theta_2 = (2,3,-1)^\top$;  then, $p_1\approx 0.691$ and  $p_2\approx  0.710$;
	\item[(a3)] ($\alpha_1\neq\alpha_2$, $\beta_1\neq 0$, $\beta_2=0$) Set $\theta_1 = (2,2,3)^\top$, $\theta_2 = (1,0,0)^\top$; then, $p_1\approx 0.691$ and $p_2\approx 0.731$;
	\item[(a4)] ($\alpha_1\neq\alpha_2$, $\beta_1=\beta_2\neq 0$) Set $\theta_1 = (2,2,3)^\top$, $\theta_2 = (-1,2,3)^\top$; then,  $p_1\approx 0.691$ and $p_2\approx 0.402$;
	\item[(a5)] ($\alpha_1\neq\alpha_2$, $\beta_1=\beta_2=0$) Set $\theta_1 = (2,0,0)^\top$, $\theta_2 = (-1,0,0)^\top$; then,  $p_1\approx 0.881$ and $p_2\approx 0.269$;
	\item[(a6)] ($\alpha_1=\alpha_2$, $\beta_1=\beta_2\neq 0$) Set $\theta_1=\theta_2 = (2,3,2)^\top$; then,  $p_1=p_2\approx 0.691$.
\end{enumerate}
In Table \ref{table_a1} and \ref{table_a2}, we present
the bias of the estimates of the regression parameters and $p_k$,  the empirical standard errors (SE), the average of the estimated standard errors (ESE) and the $95\%$ coverage probabilities (CP) with our proposed method. It is seen that for all cases, $\beta$ is always identifiable. For the identifiability of the intercepts, when the slope parameters are different across two models, $\alpha_1$ and $\alpha_2$ are both identifiable for
Cases (a1)--(a2) in Table \ref{table_a1}.  We have to emphasise that in Case (a3),   $\alpha_1$ is  identifiable but $\alpha_2$ is not identifiable with the proposed method.
On the other hand, in the case that the two subpopulations share the same $\beta$,  $\alpha_1$ and $\alpha_2$ are identifiable in Case (a4), but $\alpha_1$ and $\alpha_2$ are not identifiable in Cases (a5)--(a6). In addition, the results in Table \ref{table_a1} and \ref{table_a2} show that the proposed pooled case-control studies method produces accurate estimation for the population percentages of cases and controls.
All observations here are consistent with the discussions in section 2.1.

\subsection{Simulated data: comparing three different methods }
In the second part, we conduct simulations to examine the efficiency gain in estimating the slope parameters with our proposed method by comparing the variance of three estimators: the maximum likelihood estimator for single case-control study with known $f(\cdot)$ denoted by $\hat{\theta}^{(1)}$, the maximum likelihood estimator for single case-control study with unknown $f(\cdot)$ (Scott and Wild, 1986) denoted by $\hat{\theta}^{(2)}$ and our proposed estimator denoted by $\hat{\theta}^{(3)}$. The settings of cases (b1)--(b4) are given in Table \ref{setting_b}. For the convenience of comparison, we set the dimension of $\beta$ to be one in all cases in this subsection.
The results of cases (b1)--(b4) summarized in Tables \ref{table_b1}--\ref{table_b2} are based on  1000 replications.

By comparing the ESEs of $\beta_i$ in Tables \ref{table_b1}--\ref{table_b2}, one can observe that $\hat{\beta}^{(1)}$ is most efficient and $\hat{\beta}^{(2)}$ is least efficient among the three estimators, which confirms our theory.  Significant efficiency gain in estimating the slope parameter with our proposed method are observed when some $r_k$ is getting close to 0, as in the second case-control study in Cases (b2) and (b3).
Furthermore, as shown in Table \ref{table_b1} Case (b3),  $\hat{\beta_2}^{(1)}$ is nearly as efficient as $\hat{\beta_2}^{(3)}$ and the estimated standard errors of
$\hat{\alpha}_2^{(1)}$ are significantly smaller,
when the sample sizes of the other two case-control studies are large enough.
On the other hand, when the number of case-control studies increases,
it can be seen from Cases (b1) and (b4) in Table \ref{table_b2} that
$\hat{\beta}_i^{(3)}$ is more efficient than $\hat{\beta}_i^{(2)}$.
Lastly, we observe that when the proportion of cases $q_i$ in single case-control study sample is close to $1$ or $0$ and $|p_i-q_i|$ is large, as in the  1st, 2nd, 4th and 5th case-control studies in Case (b4), combining several case-control studies can balance the proportions between cases and controls, and thus leads to more efficient estimation for $\beta_i$.

\subsection{Hepatitis C infection dataset}
The proposed method is applied to real-life data
from a published study on liver fibrosis and cirrhosis in
patients with chronic hepatitis C infection, which is available  at https:// archive.ics.uci.edu/ml/datasets/HCV+data (Lichtinghagen {\it et al.}    2013).
It is known that progressive fibrosis is a major cause of morbidity and mortality in chronic liver disease. 
Factors having high correlations with fibrosis stages in chronic liver disease are included in this study. 		

Following the data pre-processing procedure in Hoffmann {\it et al.} (2018), $589$ patients  are selected aged between 23 to 77. They have been divided into 4 categories already: the healthy ones, the Hepatitis, the Fibrosis and the Cirrhosis, with sample size  $533$, $20$, $12$ and $24$ respectively. 
Among the 12 predictors (age, sex and $10$ other measurands),  according to Hoffmann {\it et al.} (2018), 6 predictors including ALB, BIL, CHE, GGT, AST, ALT are used in the decision tree for clustering and performs well. Hence,  
we  fit model (\ref{klg}) with the $6$ factors. 
To illustrate the idea of combining case-control studies, we randomly partition the $533$ healthy samples into three subsets with size $177$, $177$ and $179$ respectively, and obtain $3$ case-control studies on Hepatitis,  Fibrosis and Cirrhosis  as the mixture of the healthy subgroup (controls) with each diseased groups (cases) respectively.
Especially, 
the case-control study on Hepatitis consists of 177 controls and 20 Hepatitis cases; the case-control study on  Fibrosis
consists of 177 controls and 12 Fibrosis cases;   the case-control studies on Cirrhosis consists of 
179 controls and 24 Cirrhosis cases. 

We apply our combining case-control studies method to  fit model (\ref{klg}) with  $6$ predictors and $K=3$. 
We repeat our proposed method by randomly splitting the healthy group for $50$ times.  For comparison, the odds ratio method by Prentice and Pyke (1979) is applied to each  case-control study.
The results are presented in Table \ref{realdata}, 
from which one can see   that  each factor contributes to the severity of the disease.   
By combining the three case-control studies, our proposed estimator gives stable and more efficient estimates compared with the classical odds ratio estimator.


\section*{Supplementary material}
The supplementary material contains lemmas and technical proofs for the main theorems.

	\newpage
	\begin{table}[H]
	\caption{Estimation results for $\beta_1\neq\beta_2$ and $N=500$.}
	\begin{center}
		\scalebox{0.7}{
			\begingroup
			\setlength{\tabcolsep}{6pt} 
			\renewcommand{\arraystretch}{0.9} 
			\begin{tabular}{lrrrrrrrrrrrrrrrrr}
				\toprule
				Par. &\multicolumn{5}{c}{Case (a1)}&& \multicolumn{5}{c}{Case (a2)} && \multicolumn{5}{c}{Case (a3)}\\
				\cmidrule{2-6}  \cmidrule{8-12} \cmidrule{14-18}
				&True& Bias & SE & ESE & CP &&True& Bias & SE & ESE & CP &&True& Bias & SE & ESE & CP \\
				\midrule
				$\alpha_1$ & 2 & 0.060 & 0.472 & 0.467 & 0.958  && 2 & 0.040 & 0.373 & 0.367 & 0.943 && 2& $-$0.159 & 0.371 & 0.391 & 0.941\\
				$\beta_{11}$ & 2 & 0.065 & 0.329 & 0.320 & 0.946 && 2 & 0.067 & 0.343 & 0.321 & 0.938 && 2 & 0.096 & 0.329 & 0.326 & 0.964 \\
				$\beta_{12}$ & 3 & 0.111 & 0.433 & 0.423 & 0.961 && 3 & 0.106 & 0.445 & 0.423 & 0.950 && 3 &  0.150 & 0.431 & 0.429 & 0.957 \\
				$\alpha_2$ & $-$1& $-$0.039 & 0.430 & 0.442 & 0.963 && 2 & 0.035 & 0.369 & 0.362 & 0.943 && 1 & $-$16.605 & 7.738 & * & 1.000 \\ 
				$\beta_{21}$ & 3 & 0.108 & 0.447 & 0.424 & 0.951 && 3 & 0.110 & 0.412 & 0.402 & 0.956 && 0 & 0.073 & 0.146 & 0.115 & 0.844 \\
				$\beta_{22}$ & 2 & 0.075 & 0.331 & 0.322 & 0.954 && $-1$ & $-$0.033 & 0.236 & 0.228 & 0.946 && 0 & 0.106 & 0.130 & 0.120 & 0.848 \\
				\rule{-3pt}{3.5ex}
				$p_1$ & 0.691 & $-$0.004 & 0.082 & &  && 0.691 & $-$0.002 & 0.049 &  &  && 0.691 & $-$0.046 & 0.049 &  &  \\
				$p_2$ & 0.402 & $-$0.001  & 0.086 &  &  && 0.710 & $-$0.003 & 0.050 &  &  && 0.731 & $-$0.661 & 0.157 &  &  \\
				\bottomrule
			\end{tabular}
			\endgroup
	}\end{center}\footnotesize{Notes: ``Par." stands for parameter; ``True" means the true value of the parameter; the symbol * means the value is greater than $10^4$.}
	\label{table_a1}
\end{table}

	\begin{table}[H]
	\caption{Estimation results for $\beta_1=\beta_2$ and $N=500$.}
	\begin{center}
		\scalebox{0.7}{
			\begingroup
			\setlength{\tabcolsep}{6pt} 
			\renewcommand{\arraystretch}{0.9} 
			\begin{tabular}{lrrrrrrrrrrrrrrrrr}
				\toprule
				Par. &\multicolumn{5}{c}{Case (a4)}&& \multicolumn{5}{c}{Case (a5)} && \multicolumn{5}{c}{Case (a6)}\\
				\cmidrule{2-6}  \cmidrule{8-12} \cmidrule{14-18}
				&True& Bias & SE & ESE & CP &&True& Bias & SE & ESE & CP &&True& Bias & SE & ESE & CP \\
				\midrule
				$\alpha_1$ & 2 & 0.073 & 0.487 & 0.498 & 0.959 && 2 &  $-$8.343 & 10.358 & * & 0.983 &&  2 & $-$1.872 & 2.704 & 2.016 & 0.798 \\
				$\beta_{11}$ & 2 &  0.071 & 0.341 & 0.321 & 0.946 && 0 & 0.004 & 0.149 & 0.125 & 0.915  && 3 & 0.130 & 0.478 & 0.427 & 0.933 \\
				$\beta_{12}$ & 3 & 0.111 & 0.445 & 0.423 & 0.947 && 0 & $-$0.000 & 0.154 & 0.125 & 0.905 && 2 & 0.096 & 0.362 & 0.323 & 0.931 \\
				$\alpha_2$ & $-$1& $-$0.022 & 0.464 & 0.477 & 0.957 && $-$1 &  $-$6.033 & 18.097 & * & 1.000 && 2 & $-$1.828 & 2.301 & 1.917 & 0.795 \\
				$\beta_{21}$ & 2 & 0.058 & 0.319 & 0.319 & 0.959 && 0  & 0.004 & 0.146 & 0.125 & 0.924 && 3 & 0.121 & 0.451 & 0.425 & 0.956\\
				$\beta_{22}$ & 3 & 0.090 & 0.430 & 0.420 & 0.947 && 0  & 0.004 & 0.148 & 0.125 & 0.920 && 2 & 0.092 & 0.341 & 0.322 & 0.946 \\
				\rule{-3pt}{3.5ex}
				$p_1$ & 0.691 & 0.000 & 0.088 &  & && 0.881 & $-$0.608 & 0.250  & & && 0.691 & $-$0.320 & 0.274 & & \\
				$p_2$ & 0.402 & 0.003 & 0.094 & & && 0.269 & 0.010 & 0.254  &  & && 0.691 & $-$0.322 & 0.272 &  &  \\
				\bottomrule
			\end{tabular}
			\endgroup
	}\end{center}\footnotesize{Notes: ``Par." stands for parameter; ``True" means the true value of the parameter; The symbol * means the value is greater than $10^4$.}
	\label{table_a2}
\end{table}

\begin{table}[H]
	\caption{Simulation settings for Cases (b1)--(b4). }
	\begin{center}
		\scalebox{0.8}{
			\begingroup
			\setlength{\tabcolsep}{6pt} 
			\renewcommand{\arraystretch}{0.8} 
			\begin{tabular}{lrrrrrrrrrrr}
				\toprule
				Case  & $K$ &$N$&CC Study $i$& $\alpha_i$ &$\beta_i$ & $n_{i0}$& $n_{i1}$ &$n_i$&$r_i$& $q_i$ &$p_i$\\
				\midrule
				(b1)&2&830&1& $-$3& 2& 500& 10& 510& 0.615& 0.020& 0.130\\
				&&&2& $-$2& 3&  20& 300&  320& 0.277&  0.938&  0.283\\
				\rule{-3pt}{3.5ex}
				(b2)&3&1020&1& $-$3& 2& 500& 10& 510& 0.500& 0.020& 0.130\\
				&&&2& $-$2& 3&  20& 300&  320& 0.314&  0.938&  0.283\\
				&&&3& $-$1& 1&  100& 90&  190& 0.186& 0.474&  0.303\\
				\rule{-3pt}{3.5ex}
				(b3)&3&7320&1& $-$3& 2& 5000& 100& 5100& 0.697& 0.020& 0.130\\
				&&&2& $-$2& 3&  20& 300&  320& 0.043&  0.938&  0.283\\
				&&&3& $-$1& 1&  1000& 900&  1900& 0.260& 0.474&  0.303\\
				\rule{-3pt}{3.5ex}
				(b4)&5&1480&1& $-$3& 2& 500& 10& 510&  0.345 & 0.020& 0.130\\
				&&&2& $-$2& 3&  20& 300&  320& 0.216&   0.938& 0.283\\
				&&&3& $-$1& 1&  100& 90&  190& 0.128& 0.474&  0.303\\
				&&&4& 1& 2&  200& 20&  220& 0.149& 0.091&  0.648 \\
				&&&5& 4& $-$5&  200& 40&  240& 0.162& 0.167& 0.774\\
				\bottomrule
			\end{tabular}
			\endgroup
	}\end{center}\footnotesize{Notes: CC study $i$ represents the $i$-th case control study.}
	\label{setting_b}
\end{table}

\begin{table}[H]
	\caption{Estimation results for Cases (b2)-(b3).  }
	\begin{center}
		\scalebox{0.7}{
			\begingroup
			\setlength{\tabcolsep}{6pt} 
			\renewcommand{\arraystretch}{0.9} 
			\begin{tabular}{lcrrrrrrrrrrrrrrrrr}
				\toprule
				\rule{-3pt}{3.5ex}
				Case &Par. &True&&\multicolumn{4}{c}{$\hat{\theta}^{(3)}$}&& \multicolumn{4}{c}{$\hat{\theta}^{(2)}$}& &\multicolumn{4}{c}{$\hat{\theta}^{(1)}$} \\
				& &&& \multicolumn{4}{c}{ (Combining multiple CC)}&& \multicolumn{4}{c}{ (single CC with unknown $f$)}& &\multicolumn{4}{c}{(single CC with known $f$)} \\
				\cmidrule{5-8}	\cmidrule{10-13} 	\cmidrule{15-18}
				& &&& Bias & SE & ESE & CP && Bias & SE & ESE & CP  && Bias & SE & ESE & CP \\
				\midrule
				(b2)&$\alpha_1$ & $-$3&& $-$0.135 & 1.530 & 1.510 & 0.939 && $-$2.149 & 0.581 & 0.655 & 0.000 && $-$0.212 & 1.574 & 0.705 & 0.933 \\
				&$\beta_1$ & 2 && 0.096 & 0.501 & 0.484 & 0.961 && 0.098 & 0.498 & 0.500 & 0.985 && 0.087 & 0.449 & 0.430 & 0.944 \\
				&$\alpha_2$ & $-$2 && 0.063 & 0.604 & 0.601 & 0.938 && 3.659 & 0.220 & 0.308 & 0.000 && $-$0.011 & 0.204 & 0.199 & 0.940 \\
				&$\beta_2$& 3  && 0.082 & 0.467 & 0.440 & 0.957 && 0.150 & 0.619 & 0.570 & 0.966 && 0.040 & 0.358 & 0.344 & 0.937 \\
				&$\alpha_3$ & $-$1 && 0.106 & 1.071 & 1.282 & 0.977 && 0.721 & 0.078 & 0.165 & 0.000 && $-$0.015 & 0.432 & 0.442 & 0.954 \\
				&$\beta_3$& 1 && 0.032 & 0.192 & 0.194 & 0.967 && 0.030 & 0.193 & 0.194 & 0.965 && 0.031 & 0.192 &0.197 & 0.951 \\
				\rule{-3pt}{4ex}
				(b3)&$\alpha_1$& $-$3 && $-$0.043 & 0.498 & 0.494 & 0.956 && $-$2.018 & 0.161 & 0.195 & 0.000 && $-$0.006 & 0.215 & 0.219 & 0.951 \\
				&$\beta_1$ &2 && 0.008 & 0.143 & 0.148 & 0.957 && 0.007 & 0.144 & 0.150 & 0.956 && 0.005 & 0.133 & 0.134 & 0.949 \\
				&$\alpha_2$ & $-$2 && $-$0.025 & 0.302 & 0.306 & 0.955 && 3.642 & 0.202 & 0.305 & 0.000 && $-$0.017 & 0.200 & 0.199 & 0.946 \\
				&$\beta_2$ & 3 && 0.064 & 0.381 & 0.368 & 0.946 && 0.121 & 0.579 & 0.562 & 0.959 && 0.062 & 0.357 & 0.341 & 0.943 \\
				&$\alpha_3$ & $-$1 && $-$0.021 & 0.417 & 0.407 & 0.974 && 0.725 & 0.024 & 0.052 & 0.000 && 0.001 & 0.133 & 0.135 & 0.959 \\
				&$\beta_3$ & 1 && 0.007 & 0.060 & 0.060 & 0.953 && 0.007 & 0.060 & 0.060 & 0.955& & 0.007 &0.060 & 0.060 & 0.943 \\
				\bottomrule
			\end{tabular}
			\endgroup
	}\end{center}\footnotesize{Notes: ``Par." stands for parameter; ``True" means the true value of the parameter.}
	\label{table_b1}
\end{table}

\begin{table}[H]
	\caption{Estimation results for cases (b1) and (b4). }
	\begin{center}
		\scalebox{0.7}{
			\begingroup
			\setlength{\tabcolsep}{6pt} 
			\renewcommand{\arraystretch}{0.9} 
			\begin{tabular}{lcrrrrrrrrrrrrrrrrr}
				\toprule
				\rule{-3pt}{3.5ex}
				Case &Par. &True&&\multicolumn{4}{c}{$\hat{\theta}^{(3)}$}&& \multicolumn{4}{c}{$\hat{\theta}^{(2)}$}& &\multicolumn{4}{c}{$\hat{\theta}^{(1)}$} \\
				& &&& \multicolumn{4}{c}{ (Combining multiple CC)}&& \multicolumn{4}{c}{ (single CC with unknown $f$)}& &\multicolumn{4}{c}{(single CC with known $f$)} \\
				\cmidrule{5-8}	\cmidrule{10-13} 	\cmidrule{15-18}
				& &&& Bias & SE & ESE & CP && Bias & SE & ESE & CP  && Bias & SE & ESE & CP \\
				\midrule
				(b1)&$\alpha_{1}$ & $-$3&& $-$0.242 & 1.694 & 1.915 & 0.943 && $-$2.180 & 0.595 & 0.663 & 0.000 && $-$0.247 & 1.701 & 0.708 & 0.924 \\
				&$\beta_1$ & 2 && 0.121 & 0.500 & 0.489 & 0.957 && 0.117 & 0.510 & 0.502 & 0.974 && 0.085 & 0.461 & 0.430 & 0.941 \\
				&$\alpha_2$ & $-$2 && 0.100 & 0.789 & 0.832 & 0.946 && 3.651 & 0.213 & 0.307 & 0.000 && $-$0.009 & 0.203 & 0.199 & 0.942 \\
				&$\beta_2$ & 3 && 0.164 & 0.520 & 0.487 & 0.962 && 0.151 & 0.589 & 0.569 & 0.969 && 0.070 & 0.358 & 0.342 & 0.928 \\
				\rule{-3pt}{4ex}
				(b4)&$\alpha_1$ & $-$3 && $-$0.217 & 1.281 & 1.218 & 0.932 && $-$2.163 & 0.618 & 0.660 & 0.000 && $-$0.089 & 0.744 & 0.711 & 0.942 \\
				&$\beta_1$ & 2 && 0.119 & 0.501 & 0.482 & 0.962 && 0.112 & 0.528 & 0.503 & 0.965 && 0.090 & 0.453 & 0.431 & 0.934 \\
				&$\alpha_2$ & $-$2 && 0.006 & 0.491 & 0.476 & 0.934 && 3.657 & 0.202 & 0.307 & 0.000 && 0.005 & 0.200 & 0.199 & 0.946 \\
				&$\beta_2$ & 3 && 0.071 & 0.411 & 0.412 & 0.969 && 0.163 & 0.581 & 0.571 & 0.976 && 0.055 & 0.338 & 0.345 & 0.949 \\
				&$\alpha_3$ & $-$1 && $-$0.093 & 0.985 & 1.359 & 0.969 && 0.717 & 0.077 & 0.166 & 0.000 && $-$0.052 & 0.443 & 0.437 & 0.943 \\
				&$\beta_3$ & 1 && 0.034 & 0.188 & 0.193 & 0.972 && 0.033 & 0.188 & 0.194 & 0.970 && 0.034 & 0.188 & 0.194 & 0.955 \\
				&$\alpha_4$ & 1 && $-$0.037 & 0.347 & 0.387 & 0.965 && $-$2.935 & 0.164 & 0.288 & 0.000 && $-$0.043 & 0.272 & 0.298 & 0.974 \\
				&$\beta_4$ & 2&& 0.159 & 0.333 & 0.357 & 0.973 && 0.165 & 0.420 & 0.433 & 0.977 && 0.132 & 0.318 & 0.317 & 0.917 \\
				&$\alpha_5$ & 4 && 0.048 & 0.518 & 0.512 & 0.950 && $-$2.778 & 0.490 & 0.518 & 0.012 && 0.038 & 0.387 & 0.392 & 0.958 \\
				&$\beta_5$ & $-$5 && $-$0.105 & 0.728 & 0.744 & 0.963 && $-$0.215 & 0.976 & 0.926 & 0.963 && $-$0.080 & 0.668 & 0.666 & 0.946 \\
				\bottomrule
			\end{tabular}
			\endgroup
	}\end{center}\footnotesize{Notes: ``Par." stands for parameter; ``True" means the true value of the parameter.}
	\label{table_b2}
\end{table}

\begin{table}[H]
	\caption{The factors contribution to severity of Hepatitis C. }
	\begin{center}
		\scalebox{0.7}{
			\begingroup
			\setlength{\tabcolsep}{5pt} 
			\renewcommand{\arraystretch}{0.9} 
			\begin{tabular}{llrrrrrrrr}
				\toprule
				\rule{-3pt}{3.5ex}
				&& & Intercept & ALB &BIL &CHE & GGT & AST	& ALT 
				\\
				\midrule
				&Heptitis & CC & -2.2853& -0.0018 & 0.1279 & -0.08036 & 0.0833&  0.1824&  0.2812\\
				&& ESE 	&0.9956&0.0377 & 0.0635 & 0.0541 & 0.0433 & 0.0798 & 0.3352 \\
				&& OR&  - & 1.0562& 0.9828 &  0.2034 &  1.0469 &  3.0837 & -3.3217\\
				&& ESE  &  -  &  0.3447 & 0.2808 &  0.1875 & 0.3986 & 0.6535 & 0.2302\\
				\rule{-3pt}{3ex}
				&Fibrosis& Estimate & -1.5776 & -0.0238 & 0.0757 & -0.1153 &  0.0311  &  0.1938 & 0.4932 \\
				&& ESE &   0.6873& 0.0329 & 0.0428 & 0.0622 & 0.0255 & 0.0858 & 0.2356\\
				&& OR&  - &  1.3201&   1.5949 &   0.8936 &  0.8637 &  0.4984 & -1.0698\\
				&& ESE &  - & 1.8341 &   1.2701 & 1.6885 & 1.0885 &  1.5637& 1.1433\\
				\rule{-3pt}{3ex}
				&Cirrhosis& Estimate & -0.7060 &  -0.4429 &  0.4713& -0.5456&  0.3599& 0.4597& 0.4110\\
				&& ESE & 	 0.3067& 0.1996& 0.2076& 0.2416& 0.1607& 0.2015& 0.2641\\
				&& OR  & -  &  -0.9545&   1.2938 &   0.07654 &   0.3382 &  1.9898 & -21.5351\\
				&& ESE  &  -  & 1.2729 & 1.8339 & 1.2588 & 2.9500 & 2.5459 & 3.9564\\
				\bottomrule
			\end{tabular}
			\endgroup
	}\end{center}
	
	\footnotesize{Notes:  ``CC"  stands for the proposed method by combining case-control studies; ``ESE" represents the estimated standard errors for the parameter estimate;  ``OR" stands for the odds ratio estimator by Prentice and Pyke (1979). } 			
	\label{realdata}
\end{table}

\end{document}